\pdfoutput=1
\documentclass[aip,rsi,amsmath,amssymb, reprint]{revtex4-2}
\usepackage{placeins}
\RequirePackage{siunitx}
\DeclareUnicodeCharacter{00B0}{\degree}
\usepackage{hyperref}
\usepackage{graphicx}
\usepackage{dcolumn}
\usepackage{bm}
\usepackage{url}
\usepackage[utf8]{inputenc}
\usepackage[T1]{fontenc}
\usepackage{mathptmx}
\usepackage{etoolbox}

\makeatletter
\def\@email#1#2{%
	\endgroup
	\patchcmd{\titleblock@produce}
	{\frontmatter@RRAPformat}
	{\frontmatter@RRAPformat{\produce@RRAP{*#1\href{mailto:#2}{#2}}}\frontmatter@RRAPformat}
	{}{}
}%
\makeatother

\begin{document}

\title{autohaem: 3D printed devices for automated preparation of blood smears}

\affiliation{Department of Physics, Cambridge University, UK}
\affiliation{Department of Engineering, Cambridge University, UK}
\affiliation{Ifakara Health Institute, Bagamoyo, Tanzania}
\affiliation{Department of Physics, Bath University, UK}

\author{Samuel McDermott*}
\email{sjm263@cam.ac.uk}
\author{Jaehyeon Kim}
\affiliation{Department of Physics, Cambridge University, UK}
\author{Aikaterini Anna Leledaki}
\author{Duncan Parry}
\author{Louis Lee}
\author{Alexandre Kabla}
\affiliation{Department of Engineering, Cambridge University, UK}
\author{Catherine Mkindi}
\affiliation{Ifakara Health Institute, Bagamoyo, Tanzania}
\author{Richard Bowman}
\affiliation{Department of Physics, Bath University, UK}
\author{Pietro Cicuta}
\affiliation{Department of Physics, Cambridge University, UK}

\date{\today}

\begin{abstract}
\textbf{The process of making blood smears is common in both research and clinical settings, for investigating the health of blood cells and the presence of blood-borne parasites. It is very often carried out manually.  We focus here on smears for malaria diagnosis and research which are frequently analyzed by optical microscopy and require a high quality.   Automating the smear preparation promises to increase throughput and to improve the quality and consistency of the smears.  We present here two  devices (manual and motorized) designed to aid in the making of blood smears.   These are fully documented, open-source hardware, and an important principle was to make them easily fabricated locally anywhere.  Designs and assembly instructions are freely available under an open license. We also describe an image analysis pipeline for characterizing the quality of smears, and use it to optimize the settings and tunable parameters in the two devices. The devices perform as well as expert human operators, while not requiring a trained operator and offering potential advantages in reproducibility and standardization across facilities.}
\end{abstract}


\maketitle

\section{INTRODUCTION}

Blood smears are used in diagnosis for a variety of hematological disorders such as anemia.  They are also the preferred method of diagnosis of parasitic infections, such as malaria. Malaria is a widespread parasitic disease, affecting a large fraction of the world's population.  Our work is motivated by our experience in both research laboratories and clinical settings in lower-income countries, where blood smears are  made routinely. In large-scale blood testing centers (e.g. large hospitals in developed countries) this process is automated and high throughput, but there are relatively few of these facilities. We focus here on the challenges faced in the much more widespread situation of smears being performed by hand. Smear preparation is time consuming, repetitive and labour-intensive.  High quality and consistent blood smears are essential for analysis by optical microscopy of blood cells, and therefore to achieve accurate diagnoses or characterization for research.

The current ``gold standar'' for malaria diagnosis is by optical microscopy examination of blood
smears.~\cite{WHOWorldHealthOrganization2016} A thin film of the patients' blood is fixed onto a microscope slide and stained. The microscopist looks at the smear, counting the parasites in various fields of view.  This process is highly sensitive for clinical malaria, allowing differentiation of malaria species and parasite stages, and can be used to calculate the parasite density in the blood.~\cite{WHO2010}

Unfortunately, there is ample evidence that performance of routine microscopy in health facilities is
poor.~\cite{Kahama-Maro2011,Harchut2013,Sori2018,Ngasala2019} Health facilities preparing good blood films have been shown to be 24 times more likely to have accurate microscopy diagnosis than poor quality blood films, and well stained blood smears are 10 times more likely to have accurate microscopy diagnosis compared to poorly stained blood films.~\cite{Sori2018}  Moreover, competency in preparing the samples consistently appears to be lower than required.~\cite{Kiggundu2011,Mukadi2016,West2016}  Common issues with blood smears include greasy and dirty slides, unevenly spread films (too thick, thin, or streaky), poorly positioned blood films, too much or too little blood used, and issues with the Giemsa staining protocol, such as incorrect dye concentration or inconsistent buffer solution.

Our team is actively developing low-cost automation for the imaging and analysis of blood smears for malaria,~\cite{Collins2020} which tackles one aspect of robustness and consistency in the diagnostics.  It is clear that even the best image analysis algorithms will struggle with poor quality blood smears.  Poor quality smears cannot even be used for human diagnosis. In the best scenario, they need to be remade, delaying effective treatment.  This is also true in research settings, where poor quality smears can hamper experiments and affect repeatability.

This paper describes our work in creating a series of devices, which we call ``\textit{autohaem}''. \textit{autohaem} devices aim at enabling even non-experts to produce consistent, high quality, thin film blood smears at low cost. We introduce two devices, mechanical and electronic, both with completely open designs. They can be manufactured using 3D printers and assembled using common tools, enabling labs to build their own devices. This open hardware approach is already starting to revolutionize various aspects of research laboratories.~\cite{Baden2015,Amann2019,Collins2020} 3D printers are becoming increasingly common in the lower-income countries for prototyping and manufacturing.  \textit{autohaem} devices are fully open-source, with complete printing and assembly instructions,~\cite{2021AutohaemInstructions,2021AutohaemInstructionsb} assisting research laboratories who regularly make blood smears and so supporting research efforts in countries with lower resources.

A pipeline for automated analysis of smear quality is presented and used for device optimization. Red Blood Cells (RBCs), at the typical hematocrit for malaria research, are used as the testing media. This pipeline will also be suitable for a more systematic analysis of blood smear preparation, for example to help with training and evaluation of technicians.

\begin{figure}[!t]
	\centering
   \includegraphics[width=0.9\linewidth]{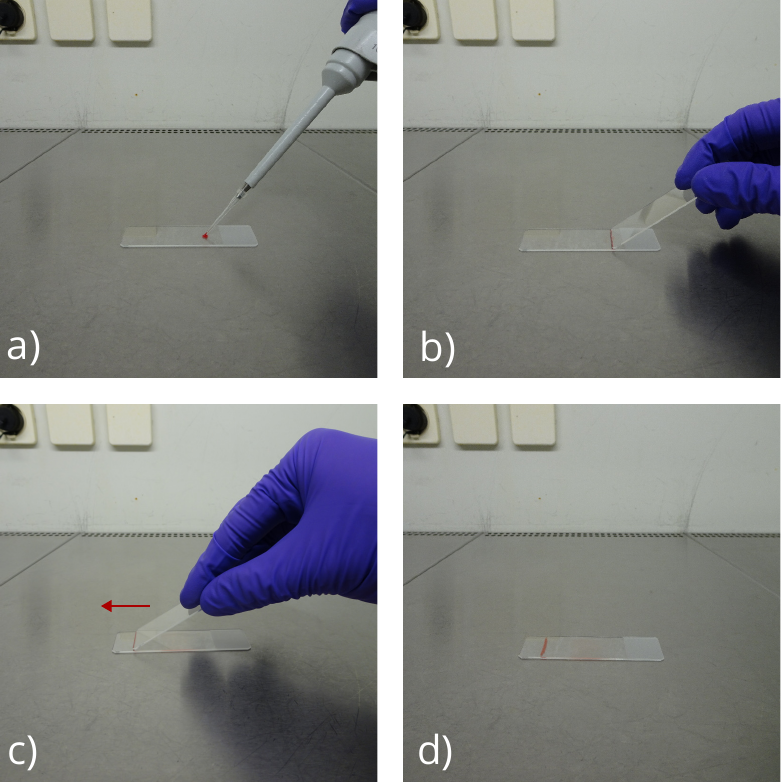}
    \caption{There are a large number of variables in the key steps of the manual smear protocol as recommended by the WHO, represented in these images.  Different users (or even the same person over time) therefore have  a high chance of performing the procedure inconsistently, leading to unusable smears and making data not comparable across labs over time. (a) Placement of a drop of blood onto a microscope slide. (b) Pulling back spreader slider into the blood drop until the blood is drawn along its edge. (c) Pushing the spreader slide along the microscope slide, at an angle of \SI{45}{\degree}, pulling the blood behind. (d) The completed blood smear.}
    \label{fig:manualsmears}
\end{figure}

\section{DESIGN PARAMETERS}
In order to achieve a sustainable, useful device, several options  have been  considered, and materials and parameters were chosen and optimized. This section describes the main design process, under the constraints of low cost and local manufacture.

\subsection{3D printing}
The devices need to be able to be 3D printed:  PLA was chosen as the material because it is a widely available and relatively cheap filament used for fused filament fabrication 3D printers. It is recyclable and biodegradable (under certain industrial conditions). Other thermoplastics are likely to perform broadly the same; stiffer materials will provide an improved rigidity to the frame, but the slide holder geometry might have to be tweaked.  If the devices become contaminated with blood, PLA can be cleaned using 70\% ethanol.

To be able to be printed with the widest range of printers possible, it is necessary to design the 3D printed parts to meet certain requirements.  All parts were designed to be printed with a layer height of \SI{150}{\micro \meter}.  None of the parts require a support material, which either require a different material such as PVA, or extra manufacturing time to clean the parts after printing.  The parts are all designed to have good adhesion to the print bed, by using curved corners and orientating the parts such that the larger surfaces are on the print bed. Therefore, they should not require ``brims'' in order to adhere to the print bed, but we recommend that the print bed should be heated. Where screws are likely to be undone, or are required to be extra strong, nut traps are used to hold nuts in place. The devices have been printed and tested on Ultimaker (S3, S5) and Prusa (i3 MK3) printers.

\begin{figure}
    \centering
    \includegraphics[width=0.9\linewidth]{2.pdf}
    \caption{The first device presented here to improve smears, the manually operated \textit{autohaem smear}. (a) 3D-rendered exploded view of \textit{autohaem smear} showing the non-3D printed parts.  (b) Photo of an \textit{autohaem smear} showing the assembled device with the two microscope slides in their positions.}
    \label{fig:autohaem-smear}
\end{figure}

\subsection{Non-printed parts}
Non-printed parts were chosen to ensure that the devices were suitable for assembly around the world.  Common machine fixings such as M3 hex head screws were used. Other non-printed parts are clearly specified in the assembly instructions, with examples of where to purchase them from.

Likewise, electronic components were specified to be purchasable from many large electronics suppliers around the world. They can be soldered using a simple hand soldering iron, and powered using common power supplies.

\begin{figure*}[!t]
    \centering
    \includegraphics[width=0.9\linewidth]{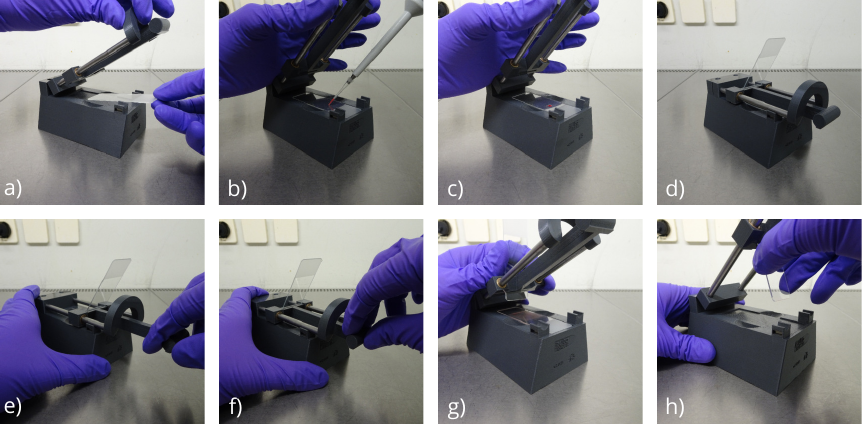}
    \caption{The \textit{autohaem smear} device makes smear production more reproducible, as characterized in this work. Once the slides and a drop of blood have been loaded, the spreader slide remains at a fixed angle and so the operator only needs to move the slider handle. The sequence of images shows the key steps in the operation of the non-motorized device. (a) Placement of the microscope slide in slot. (b) and (c) Placement of a drop of blood on a microscope slide. (d) Insertion of the spreader slide into the slider slot. (e) Pulling back slider. (f) Pushing slider forward. (g) Lifting up slider and removing the microscope slide with smear. (h) Removing the spreader slide. (Photographs are reproduced from Ref.~\cite{2021AutohaemInstructions}.)}
    \label{fig:autohaem-smear-operation}
\end{figure*}

\subsection{Spreader blade}
Most commercial automated smearing devices use single use plastic blades as a spreader.  This is not a sustainable solution, generating large amounts of plastic waste. It is also not suitable for use in devices in lower-income countries and remote testing facilities, where access to procurement of proprietary single use parts is not possible.  We designed \textit{autohaem} devices to function solely with microscope slides, which are more sustainable.

\section{CONVENTIONAL PROCESS OF SMEARING}
Blood smears are a common process in working with blood and blood diseases, and especially in lower-income countries where blood smears are made manually. The WHO's basic malaria microscopy learner's guide~\cite{HealthOrganization2010} describes the currently accepted way for making manual thin blood smears (key steps are illustrated in Fig.~\ref{fig:manualsmears}):
\begin{enumerate}
    \item Drop the blood on the slide.
    \item Using another slide as a spreader, touch the drop of blood with its edge.  The blood should be drawn along the length of the edge.
    \item Push the spreader along the slide, at an angle of \SI{45}{\degree} while remaining in constant contact with the slide. Our analysis of human experts shows that this is done at $\sim$\SI{6}{\centi \meter \per \second}.
\end{enumerate}
The steps in making blood smears for research are essentially the same, although in many  research applications, the blood product might be a refined fraction of the whole blood; for example,  only the RBCs are typically used for culturing blood stage malaria in research. For accurate optical microscopy diagnosis, which in this application implies counting parasites (and in some cases identifying their species), the thin film blood smears must have RBCs that are evenly spread.  On a macro-level, this can be seen as an absence of linear ridges or ``hesitation marks''.  In terms of RBCs density, a sweet spot needs to be achieved where cells are not overlapping, but are also not so sparse that many fields of view are required to observe a large enough cell number. The typical parasitemia for research cultures is \SI{5}{\percent}. For quantifying parasitemia for clinical purposes, the CDC recommends 500 RBCs to counted for high parasitemia ($>10\%$) and 2000 RBCs for low parasitemia ($<1\%$)~\cite{CentersforDiseaseControlandPrevention2016BloodExamination}.

\begin{figure}[!t]
	\centering
	\includegraphics[width = 0.8\linewidth]{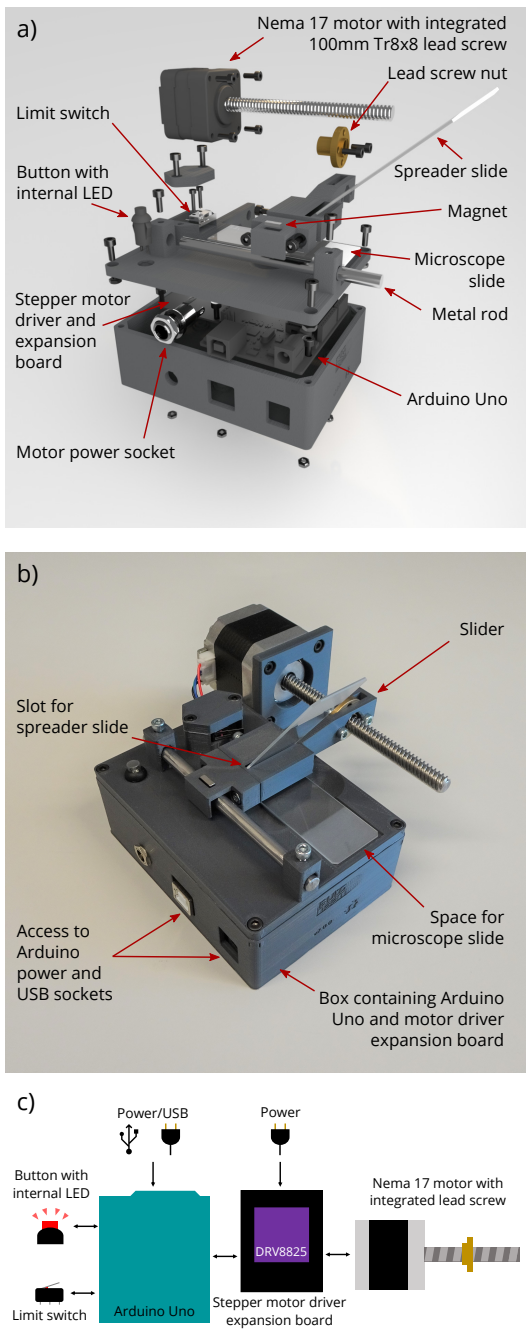}
	\caption{The second device presented in this work is motorized and improves ease of use. (a) The 3D rendered exploded view of \textit{autohaem smear+} showing the non-3D printed parts. (b) Photo of the assembled \textit{autohaem smear+}  showing the two microscope slides in their positions (Photo is reproduced from Ref.~\cite{2021AutohaemInstructionsb}). (c) Schematics of the electronics modules  of \textit{autohaem smear+}.}
	\label{fig:autohaem-smear-plus}
\end{figure}

\section{THE \textit{autohaem smear} DEVICE}
\textit{autohaem smear} is the simpler and more portable of the two devices developed here.  It is entirely mechanical and requires no electricity, making it suitable for labs or clinics with an unreliable electricity supply.  The device fixes the angle of the spreader slide,  ensures that the motion of the spreader is restricted to the horizontal axis, and that the spreader is in constant contact with the slide.  The user is able to change the speed of their smearing, but as this is the only user-changeable parameter, it is much simpler to  make reproducible and consistently high quality smears compared to a fully manual approach.

\subsection{Design}
The \textit{autohaem smear} device is shown in Fig.~\ref{fig:autohaem-smear}.  It consists of a 3D printed main body, which contains a slot to place the horizontal microscope slide.  At one end of the device, a living hinge is used to hold two parallel ground steel rods.  The hinge allows the rods to be lifted to give access to the microscope slide underneath.  The rods sit on two rod rests, with integrated magnets and  held together at the other end by a handle.  This ensures that the rods remain secured in position during smearing.  On the rods is the slider.  The slider moves smoothly along the rod with two Oilite bushings and has a handle for the user to move. The slider  contains a slit in which the spreader slide (a second microscope slide) is inserted.   The spreader slide rests on the horizontal slide with a force due to its weight.

\subsection{Operation of  \textit{autohaem smear} }
Operation of the \textit{autohaem smear} device is show in Fig.~\ref{fig:autohaem-smear-operation}.  First, the user lifts up the metal rods with the handle.  They place the microscope slide down into the slot and place the drop of blood onto the slide.  They then lower the rods into the lower position where they are held in place by the magnets. A second microscope slide, acting as the spreader, is inserted into the slot of the slider until it touches the slide underneath.  The user then pulls the handle back until the edge of the spreader meets the blood and the blood runs along the edge of the spreader.  The user then pushes the slider forward with a steady motion to produce the smear.  The slider can then be lifted, the slide with the smear removed, and the spreader slide removed through the bottom of the slider to prevent contamination of the residual blood. The edge of the spreader slide should be cleaned, and then can be used as the next microscope slide. The slide with the smear can then be fixed, stained and examined, according to the WHO protocols.~\cite{HealthOrganization2010}

\subsection{Angle of smear}
The angle that the spreader slide makes with the microscope slide is constrained by the  \textit{autohaem smear}.  By designing modifications of the slider component, it is therefore possible to adjust this angle to investigate how it affects the quality of the smear, in order to determine the optimal angle.

\section{THE \textit{autohaem smear+} DEVICE}
The second device presented here is \textit{autohaem smear+}. This is an electro-mechanical device,  suitable for laboratories or hospitals with a high throughput of blood smears. One user could prepare the next blood sample while the device is smearing, or could operate 2 devices simultaneously, dropping the blood on one, while the other one is smearing. The device is able to control the angle and speed of the spreader slide, while keeping it level and in constant contact with the microscope slide.  It is possible to program the speed of the smear (e.g., to optimize for the average environmental conditions of particular labs, which might affect the smear), and this speed should  then be set constant for every smear, regardless of the user.

\subsection{Hardware design}
The device, as shown in Fig.~\ref{fig:autohaem-smear-plus}, consists of a 3D printed main body.  This has a slot that holds the microscope slide in position, and houses the electronics. The slider is also 3D printed and, like in the \textit{autohaem smear}, holds the slide at a fixed angle.  It is attached to a lead screw nut to drive its motion on one end.  On the other end is an inverted hook with a magnet on the top.  A ground steel metal rod is held above the slide, and is used to guide the slider.  The rod prevents the slider from rotating counter-clockwise, and the magnet prevents the slider from rotating clockwise.  There is also housing for the Nema 17 motor and the limit switch.

\subsection{Electronics and Arduino code design}
The slider is driven by an Iverntech Nema 17 stepper motor with an integrated 100~mm Tr8x8 lead screw.   This was chosen because it is a common component in the  z-axis drive of 3D printers, and so it is readily available. As it is a stepper motor, its position is set precisely, enabling the slider to move at precise positions and speeds.  It is driven by a Polulu DRV8825 stepper motor driver.  This driver was chosen as it meets the high (1.5~A) current rating of the motor without additional cooling.  It also has current control, so  that a maximum current output can be set, allowing higher voltages and, therefore, higher step rates.  The driver is plugged into a stepper motor driver expansion board.  These generic boards (also usually used for 3D printers and CNC machines) provide convenient, solder-free pinouts to the driver, along with a motor connector, a decoupling capacitor, and dipswitches for microstepping.

The motor has a step size specification of \SI{1.8}{\degree} and the lead screw pitch of 8~mm, meaning that 200 steps of the motor will move the slider 8~mm.  This was micro-stepped down to 1/2 step, to increase smoothness, while being able to maintain a suitable speed (the Arduino Uno can support $\sim$4000 steps per second).  This results in the motor moving the slider 8~mm in 400 half-steps, at a maximum speed of \SI{8}{\centi \meter \per \second}. A 12~V, 2~A DC power supply is used to power the motor, connected through an external DC socket onto the motor controller expansion board.

The device is controlled using an Arduino Uno, chosen for its availability and the ease of using it to control external electronic components.  The motor driver expansion board is connected to the Arduino through the ``step'', ``direction'' and ``enable'' pins.  The Arduino is also connected to a limit switch, which defines the position of the slider at start up and an illuminated LED button, which is pressed to start the smear, and illuminates while the motor is moving.  The Arduino code makes use of the ``AccelStepper'' library~\cite{McCauley2020AccelStepper}, which provides  functions for moving the motor gracefully though acceleration and deceleration, as well as providing convenient functions for positioning and moving the motor.

\FloatBarrier

\begin{figure*}[t!]
    \centering
    \includegraphics[width=0.9\linewidth]{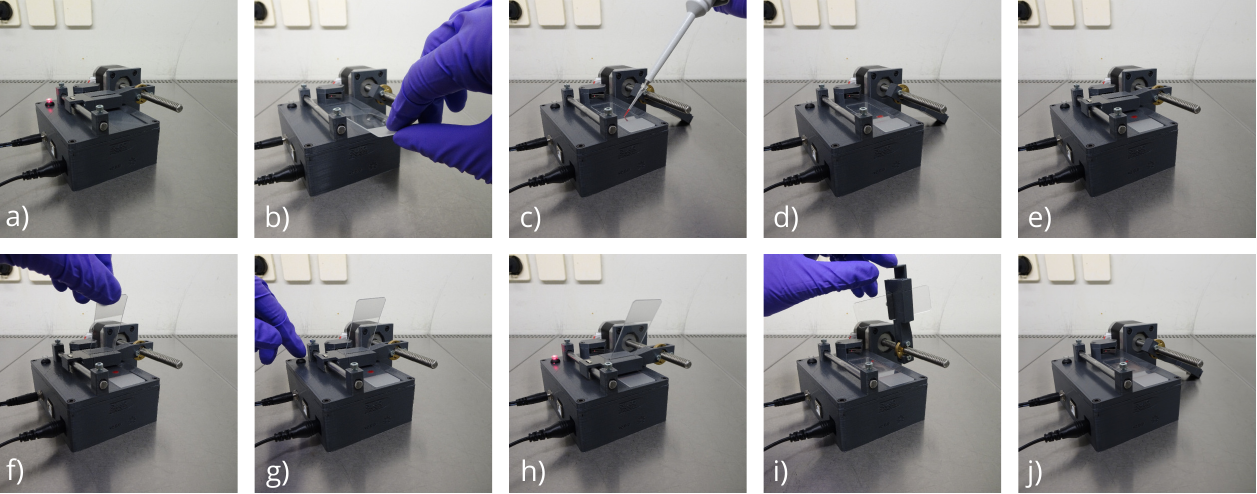}
    \caption{Operation of \textit{autohaem smear+} only requires the push of a button, once the slides and the blood drop have been loaded. The device performs very reproducibly and autonomously, so higher overall throughput can be achieved by preparing the blood while the device operates or by having multiple devices running simultaneously. The sequence of images shows the key steps in the operation of the electro-mechanical device. (a) Plugging in the device, in which the device performs calibration. (b) Rotating the slider and placing the microscope slide in the slot. (c) and d) Placing a drop of blood on the microscope slide. (e) Rotating the slider into position. (f) Inserting the spreader microscope slide. (g) Pressing the button to start the smear. (h) Device is producing the smear. (i) Removing the spreader slide. (Photos are reproduced from Ref.~\cite{2021AutohaemInstructionsb}).}
    \label{fig:autohaem-smear-plus-operation}
\end{figure*}

\subsection{Operation of \textit{autohaem smear+} }
As shown in Fig.~\ref{fig:autohaem-smear-plus-operation}, the user starts the operation of the device by plugging it in.  On startup, the device performs an initial calibration: the slider moves backward until it makes contact with the limit switch, and then moves forward to its idle position. The user then rotates the slider around the axis of the lead screw, in order to place the microscope slide in the slot.  They place their drop of blood in the center of the two circle cutouts.  The user rotates the slider back into position and inserts the spreader microscope slide into the slot in the top.  Upon pressing the button, the device starts to create the smear.  The LED illuminates to show the device is in operation and the  slider starts to move forward.  The slider then slows down as it enters the region where the blood drop is.  The slider microscope slide moves through the drop of blood, so that the blood is drawn along the edge of the spreader slide.  The slider then moves forward at a user defined speed until it reaches its idle position, where it stops.  The user rotates the slider and removes the spreader slide from underneath, preventing contamination with any residual blood. The spreader slide can then be cleaned and re-used as the microscope or spreader slider. The microscope slide is then removed. When microscope slides are in short supply, they can be cleaned and reused after examination.~\cite{HealthOrganization2010}

\subsection{Angle and speed of smear}
Two parameters can be controlled when using \textit{autohaem smear+}: the angle that the spreader slide makes with the microscope slide, and the speed of the slider as it creates the smear.  These two parameters can be adjusted to find the optimal angle and speed of the device.

\section{AVAILABILITY AND DOCUMENTATION}
Both \textit{autohaem} devices have been designed to be open source and accessible from their inception.  The 3D models have been designed in OpenSCAD, with version control using git and hosted on GitLab.~\cite{2021AutohaemSmear,2021AutohaemPlus} Code reuse between the different designs was achieved using submodules.~\cite{20217.11Submodules} The OpenSCAD files are compiled using GitLab's CI/CD tools, enabling continuous development and compilation of the STL parts. The Arduino code is also hosted on Gitlab.~\cite{2021AutohaemArduino}

Good documentation is key to enabling others to replicate designs or collaborate on future improvements.  The \textit{autohaem} device assembly instructions~\cite{2021AutohaemInstructions,2021AutohaemInstructionsb} are written to be compiled with GitBuilding,~\cite{2021Gitbuilding} allowing clear instructions that are easy to version control.  They are built using GitLab's CI/CD tools and hosted on GitLab pages, meaning there are constantly up to date instructions. They consist of a Bill of Materials, step-by-step instructions with photos for reference, electronics diagrams,~\cite{2021AutohaemDiagram} and a guide for loading the software onto the Arduino.

\section{CHARACTERIZATION OF SMEAR QUALITY AND OPTIMIZATION OF DEVICES}
In order to test the performance of the devices, they were tested using human blood.  Currently the assessment of blood smear quality is done using subjective judgment: a person is judged on their ability to make a good smear when their assessor judges that it looks correct. This is clearly a process that does not easily lead to general quality standards.   Here, we present a methodical, analytical way of judging blood smears, allowing much finer gradation of success.

\subsection{Blood smearing}
\subsubsection{Blood preparation}
\SI{200}{\micro \litre} of \SI{3}{\percent} hematocrit human RBCs in RPMI were spun down and \SI{150}{\micro \litre} supernatant was removed.  The blood cells were resuspended and \SI{3}{\micro \litre} was pipetted for each blood smear.

\subsubsection{Manual creation of smears}
Four researchers who regularly make blood smears (experts) each created five blood smears using their usual manual technique.

\subsubsection{\textit{autohaem smear} blood smears}
\textit{autohaem smear} was used by a non-expert to make five blood smears for each of these angles: [\SI{30}{\degree}, \SI{40}{\degree}, \SI{50}{\degree}, \SI{60}{\degree}, \SI{70}{\degree}]. The speed of the smear was made as consistent as possible ($\sim$\SI{5}{\centi \meter \per \second}).

\subsubsection{\textit{autohaem smear+} blood smears}
\textit{autohaem smear+} was used by a non-expert to make five blood smears for each of these angles and speeds: [\SI{30}{\degree}, \SI{40}{\degree}, \SI{50}{\degree}, \SI{60}{\degree}, \SI{70}{\degree}]; [\SI{3}{\centi \meter \per \second}, \SI{4}{\centi \meter \per \second}, \SI{5}{\centi \meter \per \second}, \SI{6}{\centi \meter \per \second}, \SI{7}{\centi \meter \per \second}, \SI{8}{\centi \meter \per \second}].

\begin{figure}[!t]
    \centering
    \includegraphics[width=0.6\linewidth]{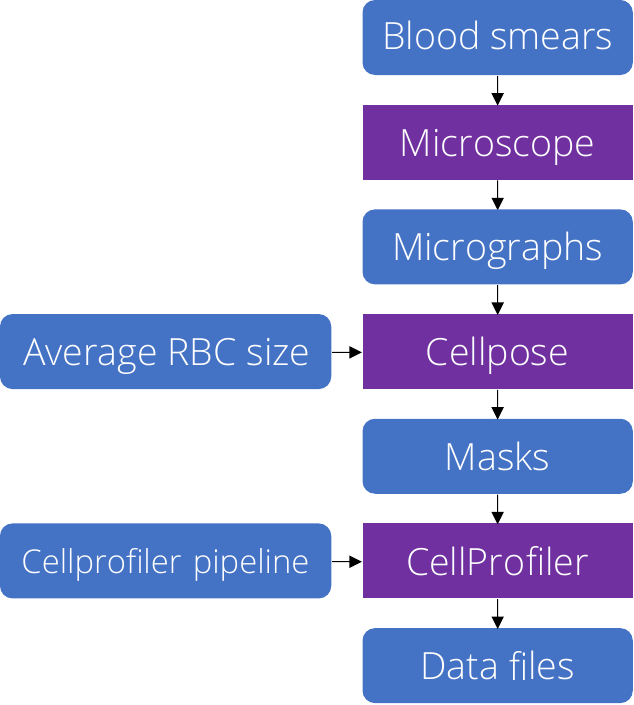}
    \caption{The smear analysis pipeline can be used to evaluate the performance of the \textit{autohaem} devices, but also potentially for teaching and evaluation of human operators.  Once the blood smears have been produced, they are imaged under a microscope producing several micrographs.  These micrographs are imaged using Cellpose, with the default model and by specifying the average RBC size in pixels. Cellpose generates a set of masks, one set for each micrograph. The set contains a number of segmented regions, each one corresponding to an individual RBC.  These masks can be used as an input for CellProfiler, with a custom CellProfiler pipeline used to extract relevant metadata about the images and RBCs to be stored in data files for further analysis.}
    \label{fig:smear_pipeline}
\end{figure}

\subsubsection{Fixing and staining}
All the smears were fixed by covering the smear with methanol for \SI{30}{\second}.  They were then stained by covering the smear in \SI{20}{\percent} Giemsa in distilled water (7.2 pH) for 10 min before rinsing and drying, similar to the standard WHO protocol~\cite{HealthOrganization2010}.

\begin{table}[t!]

	\caption{Comparison of the performance of segmentation tools for RBC smears. Eleven fields of view of manually created blood smears were segmented using each tool.  The images were also segmented by a human and the number of corrected identified (true positive), unidentified (false negative), and wrongly-identified (false positive) cells was calculated for each tool's output. From these results it was determined that Cellpose had the best precision and sensitivity.}
	\label{tab:segmentation_tools}
	\centering
	\begin{ruledtabular}
	\begin{tabular}{@{}lrr@{} }
		\multicolumn{1}{c}{Segmentation tool} & \multicolumn{1}{c}{Precision} & \multicolumn{1}{c}{Sensitivity} \\ \hline
		CellProfiler (thresholding) & 0.934 & 0.844  \\
		Weka & 0.988 & 0.917 \\
		ImageJ (Hough transform) & 0.970 & 0.948 \\
		Cellpose & 0.995 & 0.983 \\
	\end{tabular}
	\end{ruledtabular}
\end{table}

\subsection{Imaging}
All the blood smears were imaged on a Nikon Ti-E microscope using a \SI{60}{\times}, 1.4 NA objective lens. The camera was a FLIR Grasshopper 3 color camera, the images were converted to greyscale for analysis with CellPose.~\cite{Stringer2021Cellpose:Segmentation} From a starting position, 20 images were captured moving along the length of the smear ($\sim$4mm). The size of the field of view is $1920 \times 1200$ pixels ($\SI{188} \times \SI{117}{\micro \meter \squared}$).

\FloatBarrier

\subsection{Smear analysis pipeline}
\label{sec:smear_analysis_pipeline}
The smear analysis pipeline uses well documented existing tools to quantitatively evaluate the quality of thin film blood smears. The pipeline is illustrated in Fig.~\ref{fig:smear_pipeline}.  It is run as a single python notebook file.~\cite{2021SmearPipeline}

The images, converted into grayscale, are segmented using Cellpose.~\cite{Stringer2021Cellpose:Segmentation} Cellpose is a generalist deep learning-based segmentation algorithm for cells, which does not require model retraining or parameter adjustments.  Cellpose was chosen as the segmentation tool for RBCs as it had the best precision and sensitivity when tested using test smears of RBCs (Table~\ref{tab:segmentation_tools}).  The typical diameter of the cells (in pixels) is measured and used as a parameter to aid the software, and  Cellpose's default training model was used.  The software outputs a \texttt{.png} image for each original image, where each ``object'' or RBC appears as a different brightness level in the image, creating a series of masks corresponding to each RBC.  Cellpose was used to segment all the images using its python module.~\cite{Pachitariu2021Cellpose}

\begin{figure*}[p!]
    \centering
    \includegraphics[width=0.3\linewidth]{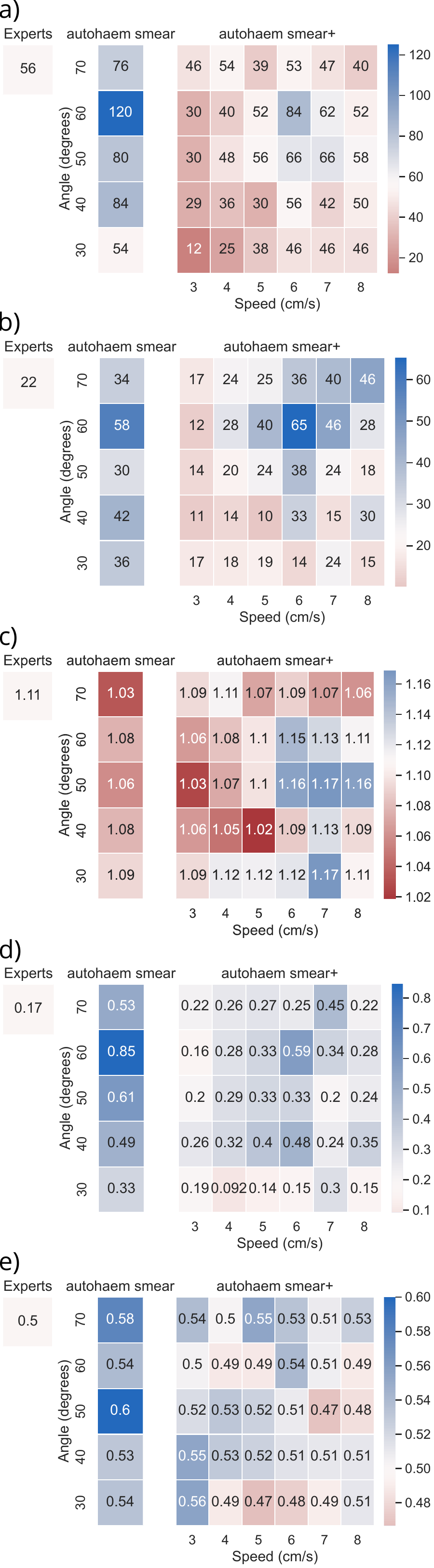}
    \caption{We optimized the two devices to the point  that they have a performance as good as, or better than, expert operators. These graphs show the comparison of results from smears created by human experts (a mean taken across the experts), vs \textit{autohaem smear}, and \textit{autohaem smear+}.  The values for the mean human expert are the center of the colormaps, meaning that regions in blue have a higher value and the regions in red have a lower value. (a) Median number of RBCs within a single field of view. (b) Interquartile range of RBCs within a single field of view. (c) Mean index of aggregation per field of view. (d) Mean adjacent neighbours per RBC. (e) Mean eccentricity per RBC.}
    \label{fig:comparison}
\end{figure*}

These masks are then processed using CellProfiler.~\cite{McQuin2018CellProfilerBiology} CellProfiler is a software tool that can create modular image analysis pipelines with many useful computational tools.  A pipeline~\cite{2021SmearPipeline} was generated using the graphical user interface (GUI), and then run using the command line.  This pipeline ingested the masks and extracted metadata based on their filenames. Any RBCs on the boundary of the image were removed.  Its modules were used to calculate the following parameters from each  image:
\begin{itemize}
    \item \textbf{RBC count:} The number of RBCs per field of view.
    \item \textbf{RBC location:} The center coordinate of each RBC.
    \item \textbf{RBC neighbors:} How many adjacent neighbours each RBC has. This includes neighbors that were previously filtered for being on the boundary of the image.
    \item \textbf{RBC eccentricity:} How circular each of the RBCs are.
\end{itemize}

Having processed the images through the smear analysis pipeline, Comma-separated values (CSV) data files were compiled containing information about the fields of view and the RBCs contained within.  Using these records, the following analyses can be carried out for each operator, device, angle and speed (these measures were averaged over the 20 fields of view per smear, for five smears):

\begin{itemize}
    \item \textbf{Median number of RBCs per field of view:} To determine the density of the smear.  Too few RBCs in a field of view mean that more fields of view are required to be examined.  Too many RBCs in a field of view mean that the image will be crowded and it will be hard to observe.
    \item \textbf{Interquartile range of RBCs per field of view:} To determine how consistent the smears are, or whether there are ranges of high and low densities.  A lower interquartile range indicates a more consistent smear.
    \item \textbf{Index of aggregation:} The Clark and Evans index of aggregation is a standard measure of clustering of a point pattern.~\cite{Clark1954DistancePopulations}  It is the ratio of the observed mean nearest neighbor distance to that expected for a Poisson distribution of the same intensity. The nearest neighbor distance for each RBC, $r_i$, (that are not touching the boundary of the field of view, but including distances to those which are) was calculated using CellProfiler. The mean distance, $\overline{r_O}$, was found for each field of view,
    \[\overline{r_O} = \frac{\sum r_i}{n}.\]
    The expected distance to the nearest neighbor, $\overline{r_E}$ was also calculated,
    \begin{eqnarray}
        \rho &=& \frac{n}{s}, \nonumber \\
        \overline{r_E} &=& \frac{1}{2\sqrt{\rho}}, \nonumber
    \end{eqnarray}
    where $n$ is the number of cells in the field of view and $s$ is the size of the field of view in pixels. The deviation of the observed pattern from the expected random pattern is measured using the ratio for index of aggregation, $R$,
    \[R = \frac{\overline{r_O}}{\overline{r_E}}.\]
    If the spatial pattern is random, $R=1$, if there is aggregation $R$ tends to 0, and if there is a regular pattern, $R$ tends to 2.15.
    \item \textbf{Adjacent neighbors:} The mean count of adjacent (touching) neighbours per RBC.  This is a measure of how clumped the RBCs are in the smear; a value of zero indicates that there are no touching or overlapping RBCs in the field of view.
    \item \textbf{RBC eccentricity:} The average eccentricity of the RBCs in a field of view. Defined by CellProfiler~\cite{BroadInstitute2020Measurement} as ``the eccentricity of the ellipse that has the same second-moments as the region. The eccentricity is the ratio of the distance between the foci of the ellipse and its major axis length.'' A RBC with an eccentricity of zero is a perfect circle, and an eccentricity of 1 is a line.  A good blood smear will have RBCs that retain their original circular shape.  If the smear is done poorly, for example, by pushing the blood rather than pulling it, the RBCs will tend to be oval.
\end{itemize}

\begin{table*}[tb!]
	\caption{The results of the nearest neighbor search to find the optimal parameters.  The expert's mean was modified to generate optimal values, which could be used in the nearest neighbor search to find the optimal parameters of the two devices.}
	\label{tab:optimal_parameters}
	\centering
	\begin{ruledtabular}
		\begin{tabular}{@{}lrrrr@{}}
			\multicolumn{1}{c}{Smear quality measurements} &
			\multicolumn{1}{c}{Experts' mean} &
			\multicolumn{1}{c}{Optimal values} &
			\multicolumn{1}{c}{\begin{tabular}[c]{@{}c@{}}\textit{autohaem smear}\\ (\SI{30}{\degree})\end{tabular}} &
			\multicolumn{1}{c}{\begin{tabular}[c]{@{}c@{}}\textit{autohaem smear+}\\ (\SI{50}{\degree}, \SI{7}{\centi \meter \per \second})\end{tabular}} \\ \hline
			Median RBC count per field of view   & 55.6   & 55.6 & 54.0   & 66.5   \\
			Interquartile range of RBC count per FoV   & 21.6   & 0  & 35.5   & 23.5   \\
			Mean index of aggregation  & 1.11 & 2.15  & 1.09 & 1.17  \\
			Mean adjacent neighbors per RBC & 0.17 & 0  & 0.331 & 0.197 \\
			Mean eccentricity per RBC  & 0.498  & 0  & 0.535 & 0.469 \\
		\end{tabular}
	\end{ruledtabular}
\end{table*}

\begin{figure*}[t!]
	\centering
	\includegraphics[width=0.9\linewidth]{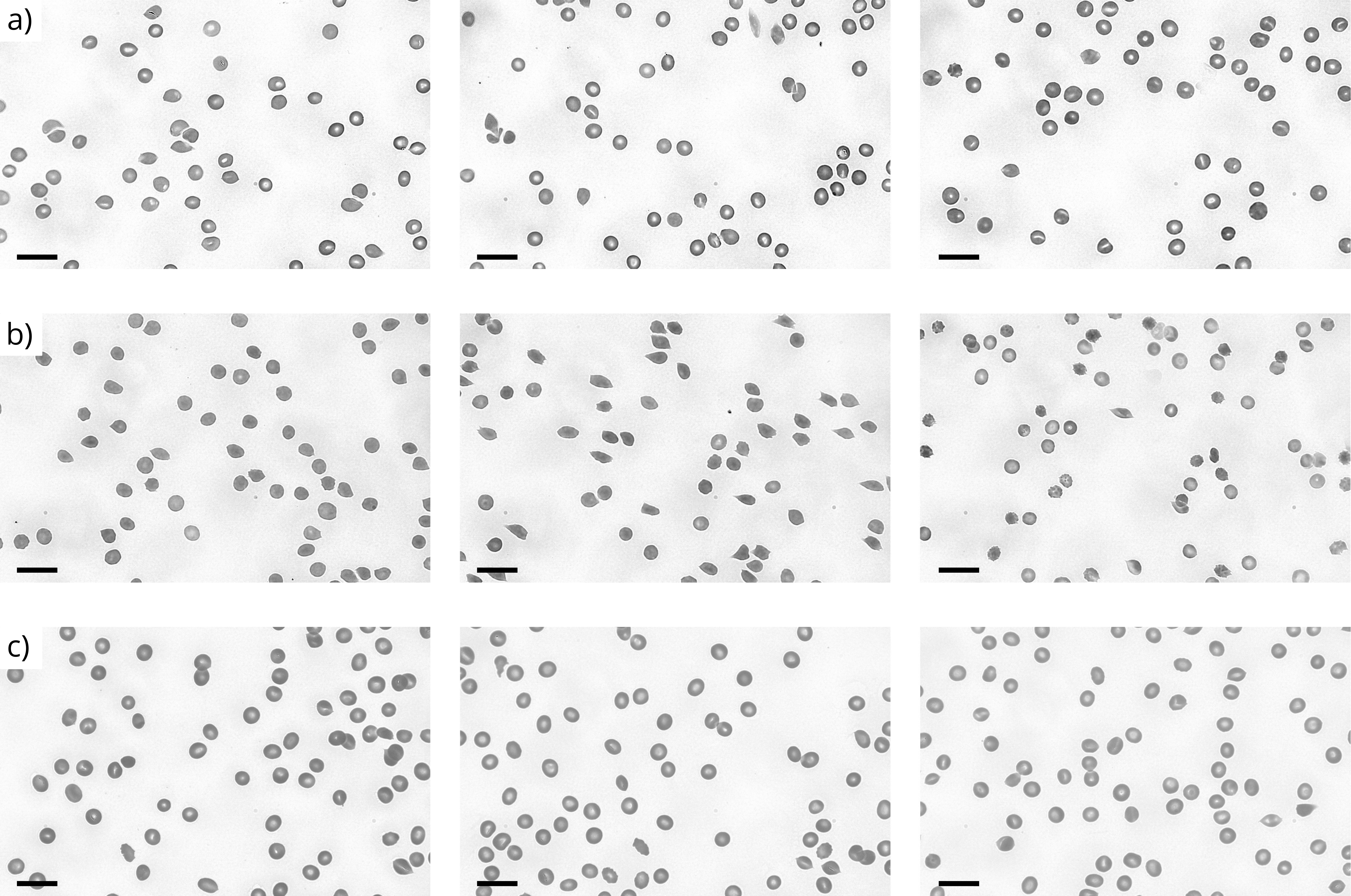}
	\caption{Typical fields of view of blood smears produced in the three conditions: (a) expert humans, (b) \textit{autohaem smear}, and (c) \textit{autohaem smear+}, in each case chosen with RBC densities close to the median density for each group. The \textit{autohaem smear} is set at the optimum angle of \SI{30}{\degree}. The density of RBCs is close to that obtained by the human experts and remains evenly distributed across the field of view (FoV). The \textit{autohaem smear+} is set at its optimum angle of \SI{50}{\degree} and speed of \SI{7}{\centi \meter \per \second}. The RBC density is higher here than that of the human experts, but there is reduced aggregation across the field of view. This performance means that a non-expert is able to use \textit{autohaem smear} and \textit{autohaem smear+} to produce a blood smear as good, or better, than that of a human expert. Images are converted to greyscale because Cellpose uses greyscale images for segmentation. Scale bars are \SI{20}{\micro \meter}.}
	\label{fig:typical-images}
\end{figure*}

\section{RESULTS}
The smear analysis was compared across the human experts, \textit{autohaem smear}, and \textit{autohaem smear+}, as shown in Fig.~\ref{fig:comparison}.  Using these  values, the optimum parameters for the devices were found.

\subsection{Device performance}
The performance of the devices was compared to the human experts, in Fig.~\ref{fig:comparison}. The colorbars are centered on the value of the mean human expert, such that squares that are blue have a higher value than the experts, and squares that are red have a lower value.

\subsubsection{Density of smear}
For \textit{autohaem smear}, increasing the angle of the spreader slide results in a higher density of RBCs [Fig.~\ref{fig:comparison}~(a)].  This reaches a maximum at \SI{60}{\degree}, with a median value of 120 RBCs per field of view.  Beyond \SI{60}{\degree}, the density decreases.  This can also be seen in the plot for \textit{autohaem smear+}, where increasing the angle corresponds to an increase in density until a maximum is reached.  A similar effect is seen for increasing speed--higher speeds produce a higher density until a maximum is reached.  This means that the highest density of RBCs is produced at \SI{60}{\degree} (the same as \textit{autohaem smear}) and \SI{6}{\centi \meter \per \second}. The mean human expert smear density is between the lowest and highest density of the devices, showing that the devices produce smears in the expected range of densities.

The reason for the reduction at high angles and speeds is because the spreader slide tends to stick-slip with these parameters.  The resulting smear consists of sharp ridges.  This effect is quantified in the interquartile range (IQR) of the density of RBCs across the fields of view [Fig.~\ref{fig:comparison}~(b)].  For \textit{autohaem smear+}, the IQR at low speeds and angles is lower, indicating that the fields of view have consistent densities of RBCs.  At higher speeds and angles, the IQR increases, showing that the ridges seen in the smears are producing inconsistent RBC densities across the fields of view.  The smears produced by moderate angles and speeds have an IQR consistent with that of the human experts.

\subsubsection{Index of aggregation}
In Fig.~\ref{fig:comparison}~(c), it can be seen that the smears which have a low speed and a medium angle have a lower index of aggregation, indicating more clumping.  At the lowest angle, \SI{30}{\degree}, the clumping is low due to the lower density of the smear.  At the higher speeds, the smear is more uniform.

\subsubsection{Adjacent neighbours}
In Fig.~\ref{fig:comparison}~(d), smears created at low angles with low speeds tend to have RBCs with fewer adjacent neighbors.  Those fields of view with  the highest density of RBCs intuitively have RBCs with more adjacent neighbors. For the fields of view with intermediate densities, there appears to be no relation between the parameters and the number of adjacent neighbors.

\subsubsection{Eccentricity}
In Fig.~\ref{fig:comparison}~(e), there appears to be no correlation between the parameters and the eccentricity of the RBCs.  The slowest and lowest angle RBCs have some of the highest eccentricities, as the blood starts to get ahead of the spreader blade and is pushed rather than pulled.

\subsection{Optimal parameters}
In order to determine the optimal parameters (angle and speed) for the devices, nearest neighbor searching was performed.  Each of the set of parameters was given a ``smear vector'', describing its performance in the five analyses.  Two such smear vector sets were created for these parameters, one for \textit{autohaem smear} and one for \textit{autohaem smear+}.

The mean of human experts was also given a smear vector, with values shown in Table~\ref{tab:optimal_parameters}.  The measurements were scaled so that they were equally important.  The optimal smear vector was also calculated by matching the median density to that of a human expert, assigning the optimal index of aggregation to 2.15 (the upper limit indicating a regular spatial pattern),  and assigning all the other measurements optimal values of 0.

A nearest neighbor search was performed between the optimal smear vector and both smear spaces. For \textit{autohaem smear}, the optimal angle was found to be \SI{30}{\degree}, with its smear vector shown in Table~\ref{tab:optimal_parameters}.  For \textit{autohaem smear+}, the optimal angle and speed were found to be \SI{50}{\degree} and \SI{7}{\centi \meter \per \second}, with its smear vector also shown in Table~\ref{tab:optimal_parameters}.  These values agree with the preconceived notions of how humans are taught to make manual smears.  They are close to the recommended \SI{45}{\degree} angle and correspond to a speed that is close to those of human experts (\SI{6}{\centi \meter \per \second}).  Typical images of smears produced by the expert humans and the two devices at their optimum parameters are shown in Fig.~\ref{fig:typical-images}.

\FloatBarrier

\section{CONCLUSIONS}
In this work, we have developed and presented the \textit{autohaem} range of devices for automated blood smearing. \textit{autohaem smear} is a mechanical device, and \textit{autohaem smear+} is an electro-mechanical device. The devices are designed to be sustainable and all the designs and assembly instructions are available under an open source license.

An automated smear analysis pipeline was created and used here as a tool for quantifying the quality of blood smears. We propose that it can have further applications in assessing human proficiency in making blood smears.

Both devices were operated exploring two tunable parameters.  Once optimized, \textit{autohaem smear} can fix the angle that a spreader slide makes with the microscope slide, and \textit{autohaem smear+} can also fix the speed that the slider moves, making smears more consistent. In this work, blood smears were made by the two devices and by human experts using the manual protocol.  The quality of the smears was quantified, and the optimum parameters of angle and speed were determined for the two \textit{autohaem} devices. The devices were shown to perform on par with human experts, even when operated by users with no experience of making blood smears.  These devices hence promise to deliver better standardization and robustness of quality across labs and over time.

The next stage of this work will be to analyze the performance of these devices using whole blood.  Improvements to the devices will include providing better protection to perform in non-ideal environments, for example, rural clinics that will be dustier than a laboratory microbiology safety cabinet, and to map out optimal device settings as a function of different temperatures and humidities. On the hardware side, a better suspension of the spreader slide would be an improvement and might reduce ridges in the smears.

\subsection*{Contributions}
S.M. designed the study, was the lead designer and maintainer of the devices, software, and assembly instructions, prepared the blood samples and reagents, performed testing of the devices, carried out the data analysis, and drafted the manuscript. J.K. contributed to the device designs, software, and assembly instructions and performed testing of the devices. A.A.L., D.P. and L.L. contributed to the device designs and data analysis. A.K. contributed to the device designs and the design of the study, and coordinated the study. C.M. contributed to the design of the study. R.W.B. contributed to the device designs and the design of the study. P.C. conceived of the study, designed the study, coordinated the study, and helped draft the manuscript. All authors gave final approval for publication and agree to be held accountable for the work performed therein.

\begin{acknowledgments}
S.M., A.K., C.M., and P.C. were funded by a GCRF QR grant awarded by Cambridge University. R.W.B. is supported by Royal Society Award No. URF\textbackslash R1\textbackslash 180153.
\end{acknowledgments}

\section*{DATA AVAILABILITY}
All materials and data are available at: \url{https://gitlab.com/autohaem}.

\bibliography{references}

\begin{thebibliography}{29}%
\makeatletter
\providecommand \@ifxundefined [1]{%
 \@ifx{#1\undefined}
}%
\providecommand \@ifnum [1]{%
 \ifnum #1\expandafter \@firstoftwo
 \else \expandafter \@secondoftwo
 \fi
}%
\providecommand \@ifx [1]{%
 \ifx #1\expandafter \@firstoftwo
 \else \expandafter \@secondoftwo
 \fi
}%
\providecommand \natexlab [1]{#1}%
\providecommand \enquote  [1]{``#1''}%
\providecommand \bibnamefont  [1]{#1}%
\providecommand \bibfnamefont [1]{#1}%
\providecommand \citenamefont [1]{#1}%
\providecommand \href@noop [0]{\@secondoftwo}%
\providecommand \href [0]{\begingroup \@sanitize@url \@href}%
\providecommand \@href[1]{\@@startlink{#1}\@@href}%
\providecommand \@@href[1]{\endgroup#1\@@endlink}%
\providecommand \@sanitize@url [0]{\catcode `\\12\catcode `\$12\catcode
  `\&12\catcode `\#12\catcode `\^12\catcode `\_12\catcode `\%12\relax}%
\providecommand \@@startlink[1]{}%
\providecommand \@@endlink[0]{}%
\providecommand \url  [0]{\begingroup\@sanitize@url \@url }%
\providecommand \@url [1]{\endgroup\@href {#1}{\urlprefix }}%
\providecommand \urlprefix  [0]{URL }%
\providecommand \Eprint [0]{\href }%
\providecommand \doibase [0]{https://doi.org/}%
\providecommand \selectlanguage [0]{\@gobble}%
\providecommand \bibinfo  [0]{\@secondoftwo}%
\providecommand \bibfield  [0]{\@secondoftwo}%
\providecommand \translation [1]{[#1]}%
\providecommand \BibitemOpen [0]{}%
\providecommand \bibitemStop [0]{}%
\providecommand \bibitemNoStop [0]{.\EOS\space}%
\providecommand \EOS [0]{\spacefactor3000\relax}%
\providecommand \BibitemShut  [1]{\csname bibitem#1\endcsname}%
\let\auto@bib@innerbib\@empty
\bibitem [{\citenamefont {{World Health
  Organization}}(2016)}]{WHOWorldHealthOrganization2016}%
  \BibitemOpen
  \bibfield  {author} {\bibinfo {author} {\bibnamefont {{World Health
  Organization}}},\ }\href@noop {} {\emph {\bibinfo {title} {{Malaria
  microscopy quality assurance manual – Ver. 2}}}}\ (\bibinfo  {publisher}
  {World Health Organization},\ \bibinfo {year} {2016})\ p.\ \bibinfo {pages}
  {140}\BibitemShut {NoStop}%
\bibitem [{\citenamefont {{WHO}}(2010)}]{WHO2010}%
  \BibitemOpen
  \bibfield  {author} {\bibinfo {author} {\bibnamefont {{WHO}}},\ }\href@noop
  {} {\emph {\bibinfo {title} {{Basic malaria microscopy: Tutor's guide}}}},\
  \bibinfo {edition} {2nd}\ ed.\ (\bibinfo  {publisher} {WHO},\ \bibinfo {year}
  {2010})\BibitemShut {NoStop}%
\bibitem [{\citenamefont {Kahama-Maro}\ \emph {et~al.}(2011)\citenamefont
  {Kahama-Maro}, \citenamefont {D'Acremont}, \citenamefont {Mtasiwa},
  \citenamefont {Genton},\ and\ \citenamefont {Lengeler}}]{Kahama-Maro2011}%
  \BibitemOpen
  \bibfield  {author} {\bibinfo {author} {\bibfnamefont {J.}~\bibnamefont
  {Kahama-Maro}}, \bibinfo {author} {\bibfnamefont {V.}~\bibnamefont
  {D'Acremont}}, \bibinfo {author} {\bibfnamefont {D.}~\bibnamefont {Mtasiwa}},
  \bibinfo {author} {\bibfnamefont {B.}~\bibnamefont {Genton}},\ and\ \bibinfo
  {author} {\bibfnamefont {C.}~\bibnamefont {Lengeler}},\ }\bibfield  {title}
  {\enquote {\bibinfo {title} {{Low quality of routine microscopy for malaria
  at different levels of the health system in Dar es Salaam}},}\ }\href
  {https://doi.org/10.1186/1475-2875-10-332} {\bibfield  {journal} {\bibinfo
  {journal} {Malaria Journal}\ }\textbf {\bibinfo {volume} {10}},\ \bibinfo
  {pages} {1--10} (\bibinfo {year} {2011})}\BibitemShut {NoStop}%
\bibitem [{\citenamefont {Harchut}\ \emph {et~al.}(2013)\citenamefont
  {Harchut}, \citenamefont {Standley}, \citenamefont {Dobson}, \citenamefont
  {Klaassen}, \citenamefont {Rambaud-Althaus}, \citenamefont {Althaus},\ and\
  \citenamefont {Nowak}}]{Harchut2013}%
  \BibitemOpen
  \bibfield  {author} {\bibinfo {author} {\bibfnamefont {K.}~\bibnamefont
  {Harchut}}, \bibinfo {author} {\bibfnamefont {C.}~\bibnamefont {Standley}},
  \bibinfo {author} {\bibfnamefont {A.}~\bibnamefont {Dobson}}, \bibinfo
  {author} {\bibfnamefont {B.}~\bibnamefont {Klaassen}}, \bibinfo {author}
  {\bibfnamefont {C.}~\bibnamefont {Rambaud-Althaus}}, \bibinfo {author}
  {\bibfnamefont {F.}~\bibnamefont {Althaus}},\ and\ \bibinfo {author}
  {\bibfnamefont {K.}~\bibnamefont {Nowak}},\ }\bibfield  {title} {\enquote
  {\bibinfo {title} {{Over-diagnosis of malaria by microscopy in the Kilombero
  Valley, Southern Tanzania: An evaluation of the utility and
  cost-effectiveness of rapid diagnostic tests}},}\ }\href
  {https://doi.org/10.1186/1475-2875-12-159} {\bibfield  {journal} {\bibinfo
  {journal} {Malaria Journal}\ }\textbf {\bibinfo {volume} {12}},\ \bibinfo
  {pages} {1--9} (\bibinfo {year} {2013})}\BibitemShut {NoStop}%
\bibitem [{\citenamefont {Sori}\ \emph {et~al.}(2018)\citenamefont {Sori},
  \citenamefont {Zewdie}, \citenamefont {Tadele},\ and\ \citenamefont
  {Samuel}}]{Sori2018}%
  \BibitemOpen
  \bibfield  {author} {\bibinfo {author} {\bibfnamefont {G.}~\bibnamefont
  {Sori}}, \bibinfo {author} {\bibfnamefont {O.}~\bibnamefont {Zewdie}},
  \bibinfo {author} {\bibfnamefont {G.}~\bibnamefont {Tadele}},\ and\ \bibinfo
  {author} {\bibfnamefont {A.}~\bibnamefont {Samuel}},\ }\bibfield  {title}
  {\enquote {\bibinfo {title} {{External quality assessment of malaria
  microscopy diagnosis in selected health facilities in Western Oromia,
  Ethiopia}},}\ }\href {https://doi.org/10.1186/s12936-018-2386-2} {\bibfield
  {journal} {\bibinfo  {journal} {Malaria Journal}\ }\textbf {\bibinfo {volume}
  {17}},\ \bibinfo {pages} {1--7} (\bibinfo {year} {2018})}\BibitemShut
  {NoStop}%
\bibitem [{\citenamefont {Ngasala}\ and\ \citenamefont
  {Bushukatale}(2019)}]{Ngasala2019}%
  \BibitemOpen
  \bibfield  {author} {\bibinfo {author} {\bibfnamefont {B.}~\bibnamefont
  {Ngasala}}\ and\ \bibinfo {author} {\bibfnamefont {S.}~\bibnamefont
  {Bushukatale}},\ }\bibfield  {title} {\enquote {\bibinfo {title} {{Evaluation
  of malaria microscopy diagnostic performance at private health facilities in
  Tanzania}},}\ }\href {https://doi.org/10.1186/s12936-019-2998-1} {\bibfield
  {journal} {\bibinfo  {journal} {Malaria Journal}\ }\textbf {\bibinfo {volume}
  {18}},\ \bibinfo {pages} {1--7} (\bibinfo {year} {2019})}\BibitemShut
  {NoStop}%
\bibitem [{\citenamefont {Kiggundu}\ \emph {et~al.}(2011)\citenamefont
  {Kiggundu}, \citenamefont {Nsobya}, \citenamefont {Kamya}, \citenamefont
  {Filler}, \citenamefont {Nasr}, \citenamefont {Dorsey},\ and\ \citenamefont
  {Yeka}}]{Kiggundu2011}%
  \BibitemOpen
  \bibfield  {author} {\bibinfo {author} {\bibfnamefont {M.}~\bibnamefont
  {Kiggundu}}, \bibinfo {author} {\bibfnamefont {S.~L.}\ \bibnamefont
  {Nsobya}}, \bibinfo {author} {\bibfnamefont {M.~R.}\ \bibnamefont {Kamya}},
  \bibinfo {author} {\bibfnamefont {S.}~\bibnamefont {Filler}}, \bibinfo
  {author} {\bibfnamefont {S.}~\bibnamefont {Nasr}}, \bibinfo {author}
  {\bibfnamefont {G.}~\bibnamefont {Dorsey}},\ and\ \bibinfo {author}
  {\bibfnamefont {A.}~\bibnamefont {Yeka}},\ }\bibfield  {title} {\enquote
  {\bibinfo {title} {{Evaluation of a comprehensive refresher training program
  in malaria microscopy covering four districts of Uganda}},}\ }\href
  {https://doi.org/10.4269/ajtmh.2011.10-0597} {\bibfield  {journal} {\bibinfo
  {journal} {American Journal of Tropical Medicine and Hygiene}\ }\textbf
  {\bibinfo {volume} {84}},\ \bibinfo {pages} {820--824} (\bibinfo {year}
  {2011})}\BibitemShut {NoStop}%
\bibitem [{\citenamefont {Mukadi}\ \emph {et~al.}(2016)\citenamefont {Mukadi},
  \citenamefont {Lejon}, \citenamefont {Barb{\'{e}}}, \citenamefont {Gillet},
  \citenamefont {Nyembo}, \citenamefont {Lukuka}, \citenamefont {Likwela},
  \citenamefont {Lumbala}, \citenamefont {Mbaruku}, \citenamefont {Veken},
  \citenamefont {Mumba}, \citenamefont {Lutumba}, \citenamefont {Muyembe},\
  and\ \citenamefont {Jacobs}}]{Mukadi2016}%
  \BibitemOpen
  \bibfield  {author} {\bibinfo {author} {\bibfnamefont {P.}~\bibnamefont
  {Mukadi}}, \bibinfo {author} {\bibfnamefont {V.}~\bibnamefont {Lejon}},
  \bibinfo {author} {\bibfnamefont {B.}~\bibnamefont {Barb{\'{e}}}}, \bibinfo
  {author} {\bibfnamefont {P.}~\bibnamefont {Gillet}}, \bibinfo {author}
  {\bibfnamefont {C.}~\bibnamefont {Nyembo}}, \bibinfo {author} {\bibfnamefont
  {A.}~\bibnamefont {Lukuka}}, \bibinfo {author} {\bibfnamefont
  {J.}~\bibnamefont {Likwela}}, \bibinfo {author} {\bibfnamefont
  {C.}~\bibnamefont {Lumbala}}, \bibinfo {author} {\bibfnamefont
  {J.}~\bibnamefont {Mbaruku}}, \bibinfo {author} {\bibfnamefont {W.~V.}\
  \bibnamefont {Veken}}, \bibinfo {author} {\bibfnamefont {D.}~\bibnamefont
  {Mumba}}, \bibinfo {author} {\bibfnamefont {P.}~\bibnamefont {Lutumba}},
  \bibinfo {author} {\bibfnamefont {J.~J.}\ \bibnamefont {Muyembe}},\ and\
  \bibinfo {author} {\bibfnamefont {J.}~\bibnamefont {Jacobs}},\ }\bibfield
  {title} {\enquote {\bibinfo {title} {{Performance of microscopy for the
  diagnosis of malaria and human African trypanosomiasis by diagnostic
  laboratories in the democratic Republic of the Congo: Results of a
  nation-wide external quality assessment}},}\ }\href
  {https://doi.org/10.1371/journal.pone.0146450} {\bibfield  {journal}
  {\bibinfo  {journal} {PLoS ONE}\ }\textbf {\bibinfo {volume} {11}},\ \bibinfo
  {pages} {1--15} (\bibinfo {year} {2016})}\BibitemShut {NoStop}%
\bibitem [{\citenamefont {West}\ \emph {et~al.}(2016)\citenamefont {West},
  \citenamefont {Gyeltshen}, \citenamefont {Dukpa}, \citenamefont {Khoshnood},
  \citenamefont {Tashi}, \citenamefont {Durante},\ and\ \citenamefont
  {Parikh}}]{West2016}%
  \BibitemOpen
  \bibfield  {author} {\bibinfo {author} {\bibfnamefont {N.}~\bibnamefont
  {West}}, \bibinfo {author} {\bibfnamefont {S.}~\bibnamefont {Gyeltshen}},
  \bibinfo {author} {\bibfnamefont {S.}~\bibnamefont {Dukpa}}, \bibinfo
  {author} {\bibfnamefont {K.}~\bibnamefont {Khoshnood}}, \bibinfo {author}
  {\bibfnamefont {S.}~\bibnamefont {Tashi}}, \bibinfo {author} {\bibfnamefont
  {A.}~\bibnamefont {Durante}},\ and\ \bibinfo {author} {\bibfnamefont
  {S.}~\bibnamefont {Parikh}},\ }\bibfield  {title} {\enquote {\bibinfo {title}
  {{An Evaluation of the National Malaria Surveillance System of Bhutan,
  2006–2012 as It Approaches the Goal of Malaria Elimination}},}\ }\href
  {https://doi.org/10.3389/fpubh.2016.00167} {\bibfield  {journal} {\bibinfo
  {journal} {Frontiers in Public Health}\ }\textbf {\bibinfo {volume} {4}},\
  \bibinfo {pages} {1--10} (\bibinfo {year} {2016})}\BibitemShut {NoStop}%
\bibitem [{\citenamefont {Collins}\ \emph {et~al.}(2020)\citenamefont
  {Collins}, \citenamefont {Knapper}, \citenamefont {Stirling}, \citenamefont
  {Mduda}, \citenamefont {Mkindi}, \citenamefont {Mayagaya}, \citenamefont
  {Mwakajinga}, \citenamefont {Nyakyi}, \citenamefont {Sanga}, \citenamefont
  {Carbery}, \citenamefont {White}, \citenamefont {Dale}, \citenamefont
  {Jieh~Lim}, \citenamefont {Baumberg}, \citenamefont {Cicuta}, \citenamefont
  {McDermott}, \citenamefont {Vodenicharski},\ and\ \citenamefont
  {Bowman}}]{Collins2020}%
  \BibitemOpen
  \bibfield  {author} {\bibinfo {author} {\bibfnamefont {J.~T.}\ \bibnamefont
  {Collins}}, \bibinfo {author} {\bibfnamefont {J.}~\bibnamefont {Knapper}},
  \bibinfo {author} {\bibfnamefont {J.}~\bibnamefont {Stirling}}, \bibinfo
  {author} {\bibfnamefont {J.}~\bibnamefont {Mduda}}, \bibinfo {author}
  {\bibfnamefont {C.}~\bibnamefont {Mkindi}}, \bibinfo {author} {\bibfnamefont
  {V.}~\bibnamefont {Mayagaya}}, \bibinfo {author} {\bibfnamefont {G.~A.}\
  \bibnamefont {Mwakajinga}}, \bibinfo {author} {\bibfnamefont {P.~T.}\
  \bibnamefont {Nyakyi}}, \bibinfo {author} {\bibfnamefont {V.~L.}\
  \bibnamefont {Sanga}}, \bibinfo {author} {\bibfnamefont {D.}~\bibnamefont
  {Carbery}}, \bibinfo {author} {\bibfnamefont {L.}~\bibnamefont {White}},
  \bibinfo {author} {\bibfnamefont {S.}~\bibnamefont {Dale}}, \bibinfo {author}
  {\bibfnamefont {Z.}~\bibnamefont {Jieh~Lim}}, \bibinfo {author}
  {\bibfnamefont {J.~J.}\ \bibnamefont {Baumberg}}, \bibinfo {author}
  {\bibfnamefont {P.}~\bibnamefont {Cicuta}}, \bibinfo {author} {\bibfnamefont
  {S.}~\bibnamefont {McDermott}}, \bibinfo {author} {\bibfnamefont
  {B.}~\bibnamefont {Vodenicharski}},\ and\ \bibinfo {author} {\bibfnamefont
  {R.}~\bibnamefont {Bowman}},\ }\bibfield  {title} {\enquote {\bibinfo {title}
  {{Robotic microscopy for everyone: the OpenFlexure microscope}},}\ }\href
  {https://doi.org/10.1364/boe.385729} {\bibfield  {journal} {\bibinfo
  {journal} {Biomedical Optics Express}\ }\textbf {\bibinfo {volume} {11}},\
  \bibinfo {pages} {2447} (\bibinfo {year} {2020})}\BibitemShut {NoStop}%
\bibitem [{\citenamefont {Baden}\ \emph {et~al.}(2015)\citenamefont {Baden},
  \citenamefont {Chagas}, \citenamefont {Gage}, \citenamefont {Marzullo},
  \citenamefont {Prieto-Godino},\ and\ \citenamefont {Euler}}]{Baden2015}%
  \BibitemOpen
  \bibfield  {author} {\bibinfo {author} {\bibfnamefont {T.}~\bibnamefont
  {Baden}}, \bibinfo {author} {\bibfnamefont {A.~M.}\ \bibnamefont {Chagas}},
  \bibinfo {author} {\bibfnamefont {G.}~\bibnamefont {Gage}}, \bibinfo {author}
  {\bibfnamefont {T.}~\bibnamefont {Marzullo}}, \bibinfo {author}
  {\bibfnamefont {L.~L.}\ \bibnamefont {Prieto-Godino}},\ and\ \bibinfo
  {author} {\bibfnamefont {T.}~\bibnamefont {Euler}},\ }\bibfield  {title}
  {\enquote {\bibinfo {title} {{Open Labware: 3-D Printing Your Own Lab
  Equipment}},}\ }\href {https://doi.org/10.1371/journal.pbio.1002086}
  {\bibfield  {journal} {\bibinfo  {journal} {PLoS Biology}\ }\textbf {\bibinfo
  {volume} {13}},\ \bibinfo {pages} {1--12} (\bibinfo {year}
  {2015})}\BibitemShut {NoStop}%
\bibitem [{\citenamefont {Amann}, \citenamefont {Witzleben},\ and\
  \citenamefont {Breuer}(2019)}]{Amann2019}%
  \BibitemOpen
  \bibfield  {author} {\bibinfo {author} {\bibfnamefont {S.}~\bibnamefont
  {Amann}}, \bibinfo {author} {\bibfnamefont {M.~v.}\ \bibnamefont
  {Witzleben}},\ and\ \bibinfo {author} {\bibfnamefont {S.}~\bibnamefont
  {Breuer}},\ }\bibfield  {title} {\enquote {\bibinfo {title} {{3D-printable
  portable open-source platform for low-cost lens-less holographic cellular
  imaging}},}\ }\href {https://doi.org/10.1038/s41598-019-47689-1} {\bibfield
  {journal} {\bibinfo  {journal} {Scientific reports}\ }\textbf {\bibinfo
  {volume} {9}},\ \bibinfo {pages} {11260} (\bibinfo {year}
  {2019})}\BibitemShut {NoStop}%
\bibitem [{202(2021{\natexlab{a}})}]{2021AutohaemInstructions}%
  \BibitemOpen
  \href@noop {} {\enquote {\bibinfo {title} {{autohaem smear v2.0.0 assembly
  instructions}},}\ }\bibinfo {howpublished}
  {\url{https://autohaem.gitlab.io/autohaem-smear/v2.0.0/}} (\bibinfo {year}
  {2021}{\natexlab{a}})\BibitemShut {NoStop}%
\bibitem [{202(2021{\natexlab{b}})}]{2021AutohaemInstructionsb}%
  \BibitemOpen
  \href@noop {} {\enquote {\bibinfo {title} {{autohaem smear+ v2.0.0 assembly
  instructions}},}\ }\bibinfo {howpublished}
  {\url{https://autohaem.gitlab.io/autohaem-smear-plus/v2.0.0/}} (\bibinfo
  {year} {2021}{\natexlab{b}})\BibitemShut {NoStop}%
\bibitem [{\citenamefont {{World Health
  Organization}}(2010)}]{HealthOrganization2010}%
  \BibitemOpen
  \bibfield  {author} {\bibinfo {author} {\bibnamefont {{World Health
  Organization}}},\ }\href@noop {} {\emph {\bibinfo {title} {{Basic Malaria
  Microscopy: Learner's Guide}}}}\ (\bibinfo  {publisher} {World Health
  Organization},\ \bibinfo {year} {2010})\ p.~\bibinfo {pages} {50}\BibitemShut
  {NoStop}%
\bibitem [{\citenamefont {{Centers for Disease Control and
  Prevention}}(2016)}]{CentersforDiseaseControlandPrevention2016BloodExamination}%
  \BibitemOpen
  \bibfield  {author} {\bibinfo {author} {\bibnamefont {{Centers for Disease
  Control and Prevention}}},\ }\href@noop {} {\enquote {\bibinfo {title}
  {{Blood Specimens - Microscopic Examination}},}\ }\bibinfo {howpublished}
  {\url{https://www.cdc.gov/dpdx/diagnosticprocedures/blood/microexam.html}}
  (\bibinfo {year} {2016})\BibitemShut {NoStop}%
\bibitem [{\citenamefont {McCauley}(2020)}]{McCauley2020AccelStepper}%
  \BibitemOpen
  \bibfield  {author} {\bibinfo {author} {\bibfnamefont {M.}~\bibnamefont
  {McCauley}},\ }\href@noop {} {\enquote {\bibinfo {title} {{AccelStepper}},}\
  }\bibinfo {howpublished}
  {\url{http://www.airspayce.com/mikem/arduino/AccelStepper/index.html}}
  (\bibinfo {year} {2020})\BibitemShut {NoStop}%
\bibitem [{202(2021{\natexlab{c}})}]{2021AutohaemSmear}%
  \BibitemOpen
  \href@noop {} {\enquote {\bibinfo {title} {{autohaem smear}},}\ }\bibinfo
  {howpublished} {\url{https://gitlab.com/autohaem/autohaem-smear}} (\bibinfo
  {year} {2021}{\natexlab{c}})\BibitemShut {NoStop}%
\bibitem [{202(2021{\natexlab{d}})}]{2021AutohaemPlus}%
  \BibitemOpen
  \href@noop {} {\enquote {\bibinfo {title} {{autohaem smear plus}},}\
  }\bibinfo {howpublished}
  {\url{https://gitlab.com/autohaem/autohaem-smear-plus}} (\bibinfo {year}
  {2021}{\natexlab{d}})\BibitemShut {NoStop}%
\bibitem [{202(2021{\natexlab{e}})}]{20217.11Submodules}%
  \BibitemOpen
  \href@noop {} {\enquote {\bibinfo {title} {{7.11 Git Tools - Submodules}},}\
  }\bibinfo {howpublished}
  {\url{https://git-scm.com/book/en/v2/Git-Tools-Submodules}} (\bibinfo {year}
  {2021}{\natexlab{e}})\BibitemShut {NoStop}%
\bibitem [{202(2021{\natexlab{f}})}]{2021AutohaemArduino}%
  \BibitemOpen
  \href@noop {} {\enquote {\bibinfo {title} {{autohaem smear plus arduino}},}\
  }\bibinfo {howpublished}
  {\url{https://gitlab.com/autohaem/autohaem-smear-plus-arduino}} (\bibinfo
  {year} {2021}{\natexlab{f}})\BibitemShut {NoStop}%
\bibitem [{202(2021{\natexlab{g}})}]{2021Gitbuilding}%
  \BibitemOpen
  \href@noop {} {\enquote {\bibinfo {title} {{gitbuilding}},}\ }\bibinfo
  {howpublished} {\url{https://gitbuilding.io}} (\bibinfo {year}
  {2021}{\natexlab{g}})\BibitemShut {NoStop}%
\bibitem [{202(2021{\natexlab{h}})}]{2021AutohaemDiagram}%
  \BibitemOpen
  \href@noop {} {\enquote {\bibinfo {title} {{autohaem smear+ electronics
  diagram}},}\ }\bibinfo {howpublished}
  {\url{https://gitlab.com/autohaem/autohaem-smear-plus/-/blob/v2.0.0/docs/images/attach_everything_together/electronics_diagram.svg}}
  (\bibinfo {year} {2021}{\natexlab{h}})\BibitemShut {NoStop}%
\bibitem [{\citenamefont {Stringer}\ \emph {et~al.}(2021)\citenamefont
  {Stringer}, \citenamefont {Wang}, \citenamefont {Michaelos},\ and\
  \citenamefont {Pachitariu}}]{Stringer2021Cellpose:Segmentation}%
  \BibitemOpen
  \bibfield  {author} {\bibinfo {author} {\bibfnamefont {C.}~\bibnamefont
  {Stringer}}, \bibinfo {author} {\bibfnamefont {T.}~\bibnamefont {Wang}},
  \bibinfo {author} {\bibfnamefont {M.}~\bibnamefont {Michaelos}},\ and\
  \bibinfo {author} {\bibfnamefont {M.}~\bibnamefont {Pachitariu}},\ }\bibfield
   {title} {\enquote {\bibinfo {title} {{Cellpose: a generalist algorithm for
  cellular segmentation}},}\ }\href
  {https://doi.org/10.1038/s41592-020-01018-x} {\bibfield  {journal} {\bibinfo
  {journal} {Nature Methods}\ }\textbf {\bibinfo {volume} {18}},\ \bibinfo
  {pages} {100--106} (\bibinfo {year} {2021})}\BibitemShut {NoStop}%
\bibitem [{202(2021{\natexlab{i}})}]{2021SmearPipeline}%
  \BibitemOpen
  \href@noop {} {\enquote {\bibinfo {title} {{smear analysis pipeline}},}\
  }\bibinfo {howpublished} {\url {https://gitlab.com/autohaem/smear-analysis}}
  (\bibinfo {year} {2021}{\natexlab{i}})\BibitemShut {NoStop}%
\bibitem [{\citenamefont {Pachitariu}\ and\ \citenamefont
  {Stringer}(2021)}]{Pachitariu2021Cellpose}%
  \BibitemOpen
  \bibfield  {author} {\bibinfo {author} {\bibfnamefont {M.}~\bibnamefont
  {Pachitariu}}\ and\ \bibinfo {author} {\bibfnamefont {C.}~\bibnamefont
  {Stringer}},\ }\href@noop {} {\enquote {\bibinfo {title} {{cellpose}},}\
  }\bibinfo {howpublished} {\url{https://pypi.org/project/cellpose/}} (\bibinfo
  {year} {2021})\BibitemShut {NoStop}%
\bibitem [{\citenamefont {McQuin}\ \emph {et~al.}(2018)\citenamefont {McQuin},
  \citenamefont {Goodman}, \citenamefont {Chernyshev}, \citenamefont
  {Kamentsky}, \citenamefont {Cimini}, \citenamefont {Karhohs}, \citenamefont
  {Doan}, \citenamefont {Ding}, \citenamefont {Rafelski}, \citenamefont
  {Thirstrup}, \citenamefont {Wiegraebe}, \citenamefont {Singh}, \citenamefont
  {Becker}, \citenamefont {Caicedo},\ and\ \citenamefont
  {Carpenter}}]{McQuin2018CellProfilerBiology}%
  \BibitemOpen
  \bibfield  {author} {\bibinfo {author} {\bibfnamefont {C.}~\bibnamefont
  {McQuin}}, \bibinfo {author} {\bibfnamefont {A.}~\bibnamefont {Goodman}},
  \bibinfo {author} {\bibfnamefont {V.}~\bibnamefont {Chernyshev}}, \bibinfo
  {author} {\bibfnamefont {L.}~\bibnamefont {Kamentsky}}, \bibinfo {author}
  {\bibfnamefont {B.~A.}\ \bibnamefont {Cimini}}, \bibinfo {author}
  {\bibfnamefont {K.~W.}\ \bibnamefont {Karhohs}}, \bibinfo {author}
  {\bibfnamefont {M.}~\bibnamefont {Doan}}, \bibinfo {author} {\bibfnamefont
  {L.}~\bibnamefont {Ding}}, \bibinfo {author} {\bibfnamefont {S.~M.}\
  \bibnamefont {Rafelski}}, \bibinfo {author} {\bibfnamefont {D.}~\bibnamefont
  {Thirstrup}}, \bibinfo {author} {\bibfnamefont {W.}~\bibnamefont
  {Wiegraebe}}, \bibinfo {author} {\bibfnamefont {S.}~\bibnamefont {Singh}},
  \bibinfo {author} {\bibfnamefont {T.}~\bibnamefont {Becker}}, \bibinfo
  {author} {\bibfnamefont {J.~C.}\ \bibnamefont {Caicedo}},\ and\ \bibinfo
  {author} {\bibfnamefont {A.~E.}\ \bibnamefont {Carpenter}},\ }\bibfield
  {title} {\enquote {\bibinfo {title} {{CellProfiler 3.0: Next-generation image
  processing for biology}},}\ }\href
  {https://doi.org/10.1371/journal.pbio.2005970} {\bibfield  {journal}
  {\bibinfo  {journal} {PLOS Biology}\ }\textbf {\bibinfo {volume} {16}},\
  \bibinfo {pages} {e2005970} (\bibinfo {year} {2018})}\BibitemShut {NoStop}%
\bibitem [{\citenamefont {Clark}\ and\ \citenamefont
  {Evans}(1954)}]{Clark1954DistancePopulations}%
  \BibitemOpen
  \bibfield  {author} {\bibinfo {author} {\bibfnamefont {P.~J.}\ \bibnamefont
  {Clark}}\ and\ \bibinfo {author} {\bibfnamefont {F.~C.}\ \bibnamefont
  {Evans}},\ }\bibfield  {title} {\enquote {\bibinfo {title} {{Distance to
  Nearest Neighbor as a Measure of Spatial Relationships in Populations}},}\
  }\href {https://doi.org/10.2307/1931034} {\bibfield  {journal} {\bibinfo
  {journal} {Ecology}\ }\textbf {\bibinfo {volume} {35}},\ \bibinfo {pages}
  {445--453} (\bibinfo {year} {1954})}\BibitemShut {NoStop}%
\bibitem [{\citenamefont {{Broad
  Institute}}(2020)}]{BroadInstitute2020Measurement}%
  \BibitemOpen
  \bibfield  {author} {\bibinfo {author} {\bibnamefont {{Broad Institute}}},\
  }\href@noop {} {\enquote {\bibinfo {title} {{Measurement}},}\ }\bibinfo
  {howpublished}
  {\url{https://cellprofiler-manual.s3.amazonaws.com/CellProfiler-4.0.5/modules/measurement.html}}
  (\bibinfo {year} {2020})\BibitemShut {NoStop}%
\end{thebibliography}%

\end{document}